\documentclass[useAMS]{mn2e}

\usepackage{graphicx,color,soul}
\usepackage{latexsym}
\usepackage{multirow}
\usepackage{amsmath,amssymb}  

\title[Modeling long GRBs using a single shock]{Modeling long GRBs using a single shock with relativistic radiation hydrodynamics}

\author[F. J. Rivera-Paleo and F. S. Guzm\'an]{F. J. Rivera-Paleo  \thanks{E-mail:friverap@ifm.umich.mx (FJRP)} and F. S. Guzm\'an \thanks{E-mail:guzman@ifm.umich.mx (FSG)}   \\ 	
        Instituto de F\'{\i}sica y Matem\'{a}ticas, Universidad
        Michoacana de San Nicol\'as de Hidalgo. \\ Edificio C3, Cd.
        Universitaria, 58040 Morelia, Michoac\'{a}n,
        M\'{e}xico.
         \\ 
        }

\begin{document}


\date{\today}

\pagerange{\pageref{firstpage}--\pageref{lastpage}} \pubyear{201X}

\maketitle

\label{firstpage}


\begin{abstract}
We explore the possibility that a single relativistic shock, where the gas dynamics is coupled with radiation, can fit the light curves of long GRBs. For this we numerically solve the one dimensional relativistic radiation hydrodynamics equations with a single initial shock. We calculate light curves due to the evolution of this shock in terms of the velocity of the shock, the opacity of the gas, mass density and density of radiated energy. We explore how the variation of each of these parameters provides different features in the light curves. As examples we include the fitting of two long GRBs.
\end{abstract}



\maketitle


\section{Introduction}
\label{sec:introduction}

Long GRBs are associated with  the  death of massive stars, creating  collimated jets moving at relativistic speeds \cite{Paczynski,Woosley}. In this scenario the radiation is so intense that it couples to the gas and the coupling contributes to the evolution of the system. In order, to study the evolution of this systems it is necessary to use models that consider the interaction between gas and radiation. The most solid theoretical model of GRBs is the fireball model, where gamma-ray photons are generated via radiative processes such as: synchrotron and inverse Compton radiation in internal or external shocks, for a review see \cite{kumar}. The internal shocks are associated to the generation of the prompt emission and the afterglow emission is associated to external shocks \cite{Rees,Dermer,Lyutikov,Kobayashi,Zhang}, although there is not a definitive model for the generation of GRBs \cite{kumar}.

In this paper we instead explore the possibility that long GRBs can be modeled with a single external shock, due to the fast propagation of a gas coupled to the radiation field. We do so by solving the equations of a relativistic gas coupled to radiation, which involves opacity and emissivity. This system is modeled with a Relativistic Radiation Hydrodynamics (RRH) system of equations, explicitly, the system of relativistic Euler plus Boltzmann transport equation for photons that we solve numerically. The source of the emission is modeled using initial conditions for a single relativistic shock. Our model involves two processes, namely the thermal radiation and  Bremsstranlung radiation with Thomson scattering. The information that leads to these mechanisms is encoded the opacity, which is one of the free parameters of our model. These two processes are considered responsible for the non-thermal optical and X-ray emission, generated by a external shock when the jet is decelerated in the surrounding medium.

The RRH system has been solved recently in various contexts. In \cite{Farris} a numerical code based on a moment formalism for radiation, under the Eddington approximation, is coupled to the MHD equations en General Relativity. In \cite{Takahashi} the radiation RMHD equations are solved in a Minkowski space-time as well. In \cite{ZanottiI} the RRH system is solved on curved background space-times in order to study the accretion onto black holes. In \cite{Fragile} the system is also solved for the accretion on black holes as well. An important condition in these models is that the opacity is restricted to be small, which eventually implies that the optical depth is also restricted to small values, because large optical depth the radiation moment equations become parabolic. Therefore, two different time scales play during the integration when solving the RRH system. On the one hand the time scale used to evolve the system related to hydrodynamics can use explicit methods and on the other hand a time scale for the radiation, which would require implicit methods. This problem has been already solved for RRH, using the IMEX method, which is a combination of explicit and implicit time evolution methods. For instance, in 
\cite{ZanottiII} and \cite{Sadowski} this method has been used to evolve accretion of gas onto a black hole using large optical depths. 

Another ingredient in models solving the RRH equations is the radiation pressure model. For instance, 
in \cite{ZanottiII} the Eddington approximation is used, which is expected to be valid in the optically thick regime, whereas in \cite{Sadowski,Takahashi} the authors use the M1 pressure approximation, which is accurate in the optically thin regime. In the context of GRBs, considering axisymmetry, thermal and no thermal processes the system has been studied by \cite{Cuesta,CuestaII,Mimica}.
 
Based on all these results, in this paper we use a 1D numerical code that solves the RRH system of equations in  two regimes of optical depth, namely the optically-thick and the optically-thin using the M1 closure scheme. We show standard shock-tube tests in two different coupling regimes between radiation and matter, that is, matter-pressure dominated and radiation-pressure dominated scenarios.

As an initial contact with observations we construct the light curves due to various combinations of parameters of given initial conditions. For this, we consider initial conditions like those used to simulate relativistic jets. The jet is constructed with the injection of gas with a given velocity and mass density onto a fluid at rest, with radiation coupled to the fluid dynamics during the process. This model allows the construction of light curves which can be compared with  the light curves of long GRBs. We explore two different scenarios: 1) in the first scenario the jets are injected with a smaller density with respect to the surrounding medium, and 2) in the second, the surrounding medium has the same density as that of the injected matter. In both scenarios the density and pressure distributions in the sorrounding medium are given by a power law.

We show a set of simulations with four free parameters: rest mass density, density of radiated energy, opacity and Lorentz factor of the injected fluid. For each of the simulations we calculate the light curve and analyze the effects of each parameter in the shape of the light curve, peak height, width and tail. In this way we can estimate the values of the four free parameters that better fit observed GRBs. As examples we present fittings of two particular cases of long GRBs in the range of 15-150KeV (GRB080413B and GRB060904B).

The paper is organized as follows. In section 2 we describe the system of equations governing the dynamics of the fluid and the numerical methods used to solve it. In section 3  we show the tests that validate our code. In section 4 we present the results for many different initial conditions of the jet and the impact of the various parameters on the light curve. We also fit as examples two long GRBs with our model. Finally, in section 5 we present final comments and conclusions.
 

\section{Relativistic radiation hydrodynamics}
\label{sec:RRH}

\subsection{Covariant evolution equations}

The fluid is modeled as an ideal fluid immersed in the radiation field with a two terms stress-energy tensor $T^{\alpha\beta}= T^{\alpha\beta}_m + T^{\alpha\beta}_r$. The first term is the stress energy tensor of a perfect fluid

\begin{equation}
T^{\alpha\beta}_m = \rho hu^\alpha u^\beta + Pg^{\alpha\beta},
\end{equation}

\noindent where $g^{\alpha\beta}$ is the metric of the spacetime, $u^\alpha$ is the four-velocity of fluid elements, $\rho$, $h=1+\epsilon+P/\rho$, $\epsilon$ and $P$ are the rest-mass density, specific enthalpy, specific internal energy and the thermal pressure, respectively. The thermal pressure is related to $\rho$ and $\epsilon$ through a gamma-law equation of state (EOS) $P=\rho\epsilon(\Gamma-1)$, where $\Gamma$ is the adiabatic index of the gas. The second term describes the radiation field and is given by

\begin{equation}
T^{\alpha\beta}_r = (E_r+P_r)u^\alpha u^\beta +F^\alpha_r u^\beta +u^\alpha F^\beta_r + P_rg^{\alpha\beta},
\end{equation}

\noindent where the zeroth, first and second radiation moments are associated with the density of radiated energy $E_r$, radiated flow $F^\alpha_r$ and radiation pressure $P_r$ respectively. The interaction between the fluid and radiation field is given through the coefficients of absorption, emission and scattering of photons. These coefficients are given by \cite{MihalasII}

\begin{equation}
G^{\alpha}_r = \frac{1}{c}\int (\chi I - \eta)n^\alpha d\nu d\Omega,  
\end{equation}

\noindent with $\eta$ is the coefficient of emissivity, $\chi$ is the coefficient of opacity and $G^\alpha_r$ is the radiation four-force density. We assume an isotropic and coherent scattering, that is, the coefficients of thermal emissivity and opacity are related through Kirchhoff's law. Thus, the fluid and radiation fields are in local thermal equilibrium, and the opacity coefficients are independent of the frequency, and they obey the ``gray-body" approximation \cite{MihalasII}. Under these conditions we can write $G_r$ in covariant form as

\begin{equation}
G^\alpha_r = \chi^t(E_r - 4\pi B)u^\alpha + (\chi^t + \chi^s)F^\alpha_r,\label{eq:Galpha}
\end{equation}

\noindent here $\chi^t =\kappa^t\rho$, $\chi^s=\kappa^s\rho$ are the opacity coefficients, where the superscript $t$ and $s$ denote the thermal and scattering opacities respectively. The $\kappa's$ are  independent of frequency, $B=\frac{1}{4\pi}a_rT^4_\text{fluid}$ is the Planck function, $T_\text{fluid}$ the temperature of the fluid and $a_r$ the radiation constant. Thus the equations governing the evolution of the system are:

\begin{eqnarray}
\nabla_\alpha(\rho u^\alpha) &=&0,\\
\nabla_\alpha T^{\alpha\beta} &=&0,\\
\nabla_\alpha T^{\alpha\beta}_r &=&-G^\beta_r.
\end{eqnarray}

\noindent For our purpose we specifically choose the Minkowski space-time in 1D and Cartesian coordinates. Finally, under all these assumptions, the RRH equations are written in the following flux balance form $\partial_t\textbf{U}+\partial_x\textbf{F}^x=\textbf{S}$, where  $\textbf{U}$ is the vector of conserved variables, $\textbf{F}^x$ are the fluxes and  $\textbf{S}$ the sources given by

\begin{equation}
 \textbf{U} = \begin{bmatrix}
      D  \\[0.3em]
      S \\[0.3em]
      \tau \\[0.3em]
      S_r\\[0.3em]
      \tau_r
     \end{bmatrix}
       = \begin{bmatrix}
       \rho W \\[0.3em]
       \rho hW^2v_x\\[0.3em]
       \rho hW^2-P-\rho W\\[0.3em]
       (E_r+P_r)W^2v_x+WF^x_r(v_x+1)\\[0.3em]
       (E_r+P_r)W^2+2Wv_xF^x_r-P_r
     \end{bmatrix},\label{eq:consvars}
\end{equation}
\begin{equation}
 \textbf{F}^x = \begin{bmatrix}
      v_xD  \\[0.3em]
       \rho hW^2v_x^2 +P\\[0.3em]
      \rho hW^2v_x - \rho Wv_x \\[0.3em]
      (E_r+P_r)W^2v_x^2 +2WF^x_rv_x +P_r\\[0.3em]
      S_r
     \end{bmatrix},
\ \ \textbf{S} =\begin{bmatrix}
       0 \\[0.3em]
       G^x_r\\[0.3em]
       G^0_r\\[0.3em]
       -G^x_r\\[0.3em]
       -G^0_r
     \end{bmatrix}, 
\end{equation}

\noindent where $W$ is the Lorentz factor, and we can write the radiation four-force $G^\alpha_r$ in terms of the conservative variables of the radiation field as \cite{ZanottiII}

\begin{eqnarray}
G^0_r &=& -W[\chi^ta_rT^4_\text{fluid} + \tau_r(2\chi^s(1-\frac{3}{1+2W^2})-\chi^t) \\
\nonumber
      & & +S_rv_x(\chi^t+\chi^s(\frac{3}{1+2W^2}-2))],\\
G^x_r &=& -\chi^ta_rT^4_\text{fluid}v_xW + \frac{\chi^t+\chi^s}{W}S_r +\tau_rWv_x\\
\nonumber
      & & [\chi^t(1-\frac{4}{1+2W^2}) + 2\chi^s(\frac{1}{1+2W^2} - 1)]+\\
\nonumber
      & & S_rv^2_xW[\chi^t(\frac{2}{1+2W^2}-1)+\chi^s(2-\frac{1}{1+2W^2})].
\end{eqnarray}

The above set of equations is closed with an EoS that relates the second moment of radiation with one of the lower order moments. The simplest approach is the Eddington approximation, which assumes a nearly isotropic radiation field and in the fluid frame shows a pressure tensor with the form \cite{MihalasII}

\begin{equation}
P^{ij}_r = \frac{1}{3}E_r\delta^{ij}.
\end{equation} 

\noindent This assumption is valid only in the optically thick regime within the diffusion limit. The radiation field in the optically thin regime requires a more general assumption. A scheme that allows a description of the radiation field in both optically thick and thin regimes is the M1 \cite{Levermore,Dubroca,Gonzalez}. The M1 closure provides a better approximation than Eddington to the radiation field, because it describes the diffusion limit, as well as the free-streaming limit where the radiative energy is transported at the speed of ligh, this closure is given by

\begin{equation}
P^{ij}_r = \left(\frac{1-\zeta}{2}\delta{ij} + \frac{3\zeta-1}{2}\frac{f^if^j}{|f|^2}\right) E_r,
\end{equation}  

\noindent where $f^i=F^i/cE$ is the reduced radiative flux and $\zeta=\frac{3+4f^if_i}{5+2\sqrt{4-3f^if_i}}$ is the Eddington factor \cite{Levermore}. We can see that this closure relation recovers the two regimes of radiative transfer. In the optically thick regime $F^i\approx 0$, $f^i=0$, and $\zeta=1/3$ which corresponds to Eddington approximation. On the other hand, in the optically thin regime $F^i= cE$, $f^i=1$, and $\zeta=1$ which corresponds to the free-streaming limit.

The gas temperature is estimated from the ideal-gas EoS via the expression $T_\text{fluid} = \frac{\mu m_p}{k_B}\frac{P}{\rho}$, being $k_B$ the Boltzmann constant, $\mu$ the mean molecular weight and $m_p$ the mass of proton. This is a good approximation when the fluid pressure is much greater than the radiation pressure. The temperature should be calculated taking into account contributions of both barions and radiation pressure. An approximate expression for the total pressure is written as

\begin{equation}
P_t = \frac{k_B}{\mu m_p}\rho T_\text{fluid} + (1 - e^{-\tau})\zeta(T) a_rT_\text{rad}^4,
\label{tem}
\end{equation}

\noindent where $\tau=\int(\chi^t+\chi^s)ds$ is the total optical depth and $T_\text{rad}=a_rE_r$ the temperature of radiation  \cite{Cuesta}. Here $\tau$ depends on the temperature only if any of the opacity coefficients does. In general the Eddington factor depends on the temperature $\zeta=\zeta(T_\text{fluid})$. When the fluid and radiation are in local thermal equilibrium (LTE), that is  $T_\text{fluid}=T_\text{rad}$, the temperature approximately obeys a fourth order equation similar to that above \cite{Cuesta}.

The inversion of the radiation variables can be written as a set of algebraic equations that can be solved independently of the primitive hydrodynamic variables:

\begin{eqnarray}
E_r &=&-3\frac{W^2}{1+2W^2}[ 2S_rv_x + \tau_r(1/W^2 - 2) ],\\ 
F^x_r &=& \frac{S_r}{W} - \frac{4}{3}E_rWv_x - F^0_rv_x,
\end{eqnarray}

\noindent where

\begin{eqnarray}
F^0_r &=& \frac{W}{1+2W^2}[ -4\tau_r(W^2-1) + \\
\nonumber
      & & (4W^2-1)S_rv_x ].
\end{eqnarray}

\noindent With all these ingredients it only remains to solve the flux balance system for the variables (\ref{eq:consvars}). The advantage of writing the RRH equations in flux balance form, is that one can use use finite volume discretization in order to apply high-resolution shock capturing (HRSC) schemes and solve the evolution problem.

\subsection{Numerical Methods}

Radiation hydrodynamics equations are difficult to integrate numerically, because they present two different time scales, a long one related to hydrodynamics and a short one associated to radiation. Using HRSC for the whole system has the disadvantage that radiation equations may develop instabilities. This becomes evident in the optically-thick regime  (diffusion limit), when the equations describing the radiation field transform from hyperbolic into parabolic, making the integration in time both inefficient and possibly unstable with an explicit method.

One way to solve this problem is to combine implicit and explicit time integrations, using an explicit scheme for the hydro and an implicit one for the radiation \cite{ZanottiII}. The combination of these two schemes has become the set of IMEX methods \cite{Uri,Pareschi}.  In the present work we use  the 2nd order IMEX Runge-Kutta which makes possible to explore larger optical depths than with an explicit RK \cite{ZanottiII}.

Following this idea, we split the right hand sides of the evolution equations for the conservative quantities  $\textbf{U}$ into two parts \cite{ZanottiII}. The vector $\textbf{f}(\textbf{U})$ will be defined for the conserved hydrodynamical variables $\{D,S,\tau\}$, and includes the respective source terms $\{0,G^x_r,G^0_r\}$, whereas $\textbf{g}(\textbf{U})$ is defined for the radiation variables $\{S_r,\tau_r\}$ including $\{-G^x_r,-G^0_r\}$ which are the terms affected by stiffness. With this in mind, the IMEX scheme takes the form

\begin{equation}
\partial_t \textbf{U} = \textbf{f}(\textbf{U}) + \textbf{g}(\textbf{U}),
\end{equation}

\noindent whose general solution with $s$ intermediate steps reads

\begin{eqnarray}
\textbf{U}^i     &=& \textbf{U}^n + dt\sum^{i-1}_{j=1}\tilde{a}_{ij}\textbf{f}(t^n + \tilde{c}_jdt,\textbf{U}^j) +\\
\nonumber
        & & dt\sum^{i}_{j=1}a_{ij}\textbf{g}(t^n + c_jdt,\textbf{U}^j), \ \ i=1...s,\\
\textbf{U}^{n+1} &=& \textbf{U}^n + dt\sum^{s}_{i=1}\tilde{b}_i\textbf{f}(t^n + \tilde{c}_idt,\textbf{U}^i) + \\
\nonumber
        & & dt\sum^{s}_{i=1}b_i\textbf{g}(t^n + c_idt,\textbf{U}^i),\\
\end{eqnarray}

\noindent where $\textbf{U}^i$ are auxiliary intermediate values of the IMEX RK scheme and $\textbf{U}^{n+1}$ is the final full step solution. The order of the method depends on the coefficients $a$, $b$, $c$, $\tilde{a}$, $\tilde{b}$ and $\tilde{c}$. The tilde coefficients are associated with the explicit part of the method, whereas those without tilde are related to the implicit part.

The IMEX scheme in this paper is a hybrid between a second order accurate one step implicit RK Gauss  \cite{Alexander}, and explicit midpoint RK that is also second order accurate. The coefficients that define the intermediate steps of this particular accuracy are given by \cite{Alexander}

\begin{equation}
\begin{array}{l| r}
      \tilde{c} & \tilde{a}_{ij}\\
       \hline
               & \tilde{b}^T
     \end{array}
\  =\begin{array}{l| c  r}
       0   & 0   & 0\\
       1/2 & 1/2 & 0\\
      \hline
           & 0   & 1
     \end{array}, 
\ \ \ \ \ \
\begin{array}{l| r}
      c & a_{ij}\\
       \hline
               & b^T
     \end{array}
\  =\begin{array}{l| c  r}
       0   & 0   & 0\\
       1/2 & 0 & 1/2\\
      \hline
           & 0   & 1
     \end{array}.
\end{equation}

\noindent Therefore, the construction of $U^{n+1}$ is explicitely

\begin{eqnarray}
U^*     &=& U^n + \frac{dt}{2}[f(t^n + \frac{1}{2}dt,U^n) + 
\nonumber\\
        & & g(t^n + \frac{1}{2}dt,U^*)],\label{eq:implicitpart}\\
U^{n+1} &=& U^n + dt[f(t^n + \frac{1}{2}dt,U^*) + 
\nonumber\\
        & & g(t^n + \frac{1}{2}dt,U^*)].\label{eq:explicitpart}
\end{eqnarray}

\noindent The implicit nature of the algorithm can be seen in the calculation of $U^*$, which appears in both sides of (\ref{eq:implicitpart}). We solve for $U^*$ using a Newton-Raphson method. In order to show that the implementation of the above algorithm works properly, we present canonical tests in the following section. 


\section{Numerical Tests}
\label{sec:NT}

There is a series of tests of the perfect fluid plus radiation system that a code has to sort out in the Eddington approximation \cite{Farris,ZanottiII,Fragile,Sadowski}. These tests are shock tube tests assuming various initial conditions shown in for instance \cite{Farris,ZanottiII}. Like in pure hydrodynamic shock tube tests, the system has initial conditions for  a gas in two different constant states (left and right), separated by a membrane. The membrane is removed at $t=0$ and the system is allowed to evolve. The initial conditions are set in Table \ref{t1}. The initial values for the radiation flux, in all cases is set to $F^x_r = 10^{-2} E_r$, in order to lie within the Eddington approximation \cite{Fragile}. The scattering opacity $\chi^s$ is set to zero in all the cases.

Each test is evolved in time until the system approaches a stationary state, except Test 5. The spatial domain $x\in[-20,20]$ is covered with $800$ equal cells. The formula used to approximate the numerical fluxes is HLLE and we use the MC slope limiter. We use the Courant-Friedrichs-Lewy factor unchanged and equal to $0.25$. 

The various tests are the following.
 
\begin{figure}
\includegraphics[width=8.5cm]{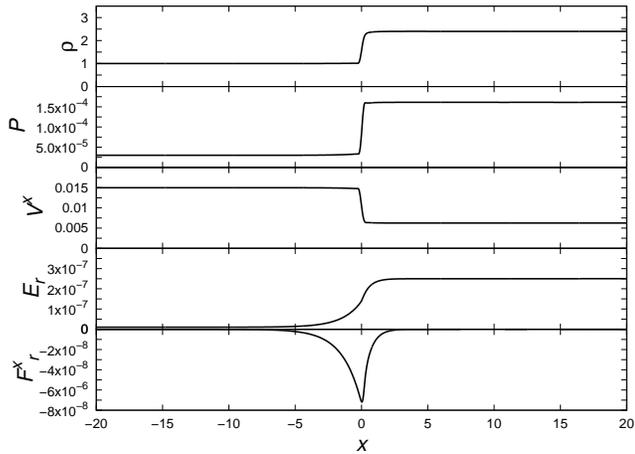}
\caption{\label{fig:t1}Solution of test 1 at $t=4000$. From top to bottom the panels show the rest-mass density, pressure, velocity, radiation energy density and radiation flux.}
\end{figure}

(i) \textit{Non-relativistic strong shock}. This is illustrated with Test 1. In this test a strong, gas-pressure dominated, but non-relativistic shock is propagating into a cold gas. The standard snapshot is shown in Figure \ref{fig:t1} and can be compared with Fig 1 of \cite{Farris}. 

\begin{figure}
\includegraphics[width=8.5cm]{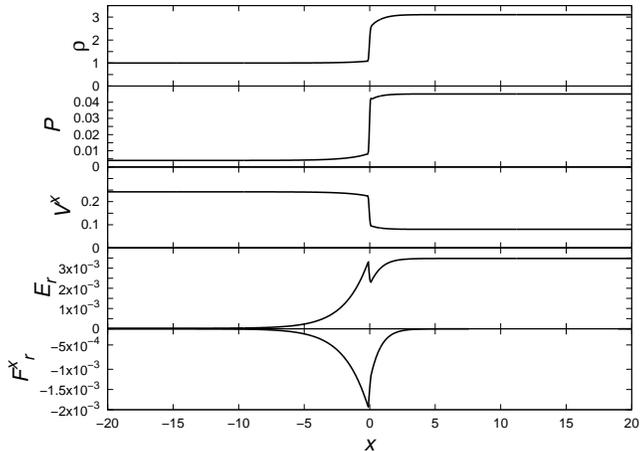}
\caption{\label{fig:t2}Solution of test 2 at $t=3000$. We show the rest-mass density, pressure, velocity, radiation energy density and radiation flux.}
\end{figure}

(ii) \textit{Mildly-relativistic strong shock}. This is Test 2 and corresponds to a strong and gas-pressure dominated shock. The radiation profiles for this case show a discontinuity in the radiation energy density and the radiative flux as shown in Figure \ref{fig:t2}. 

\begin{figure}
\includegraphics[width=8.5cm]{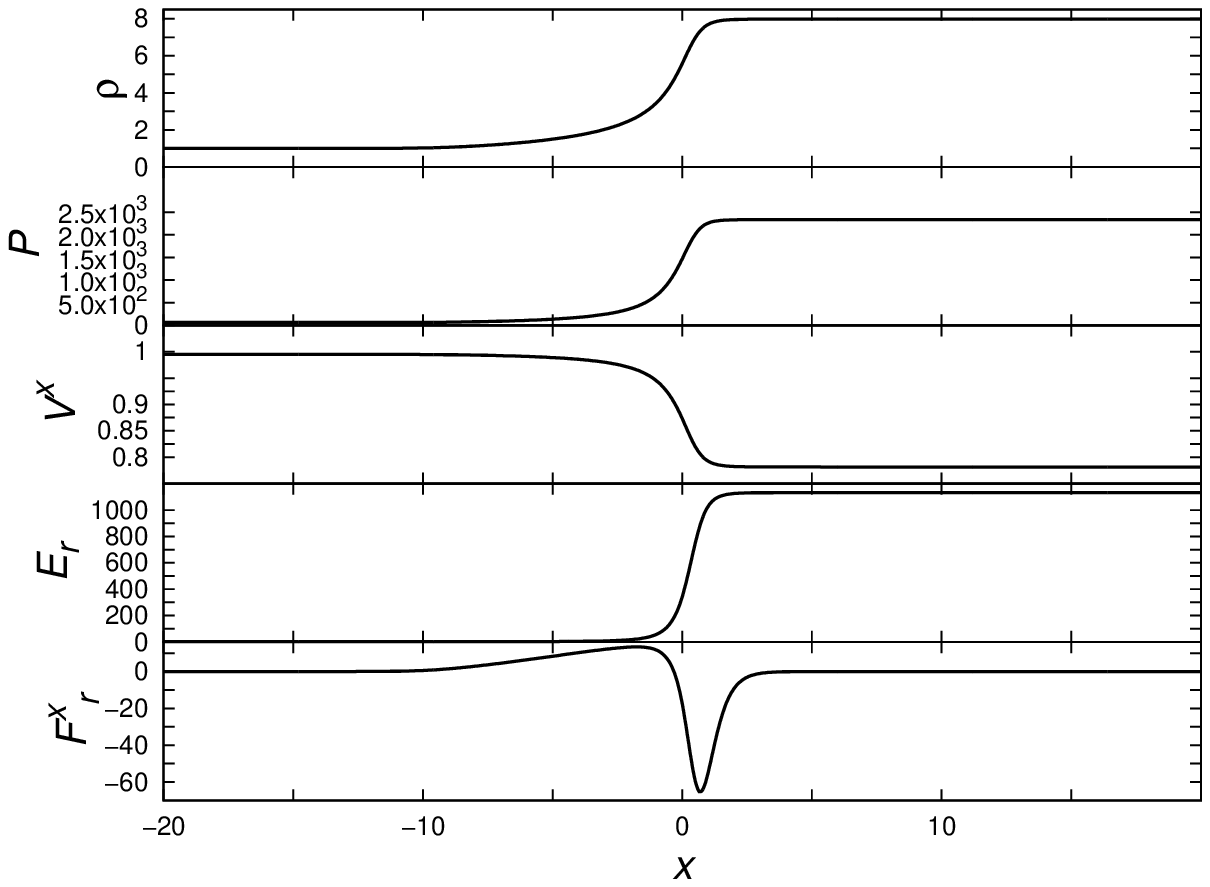}
\includegraphics[width=8.5cm]{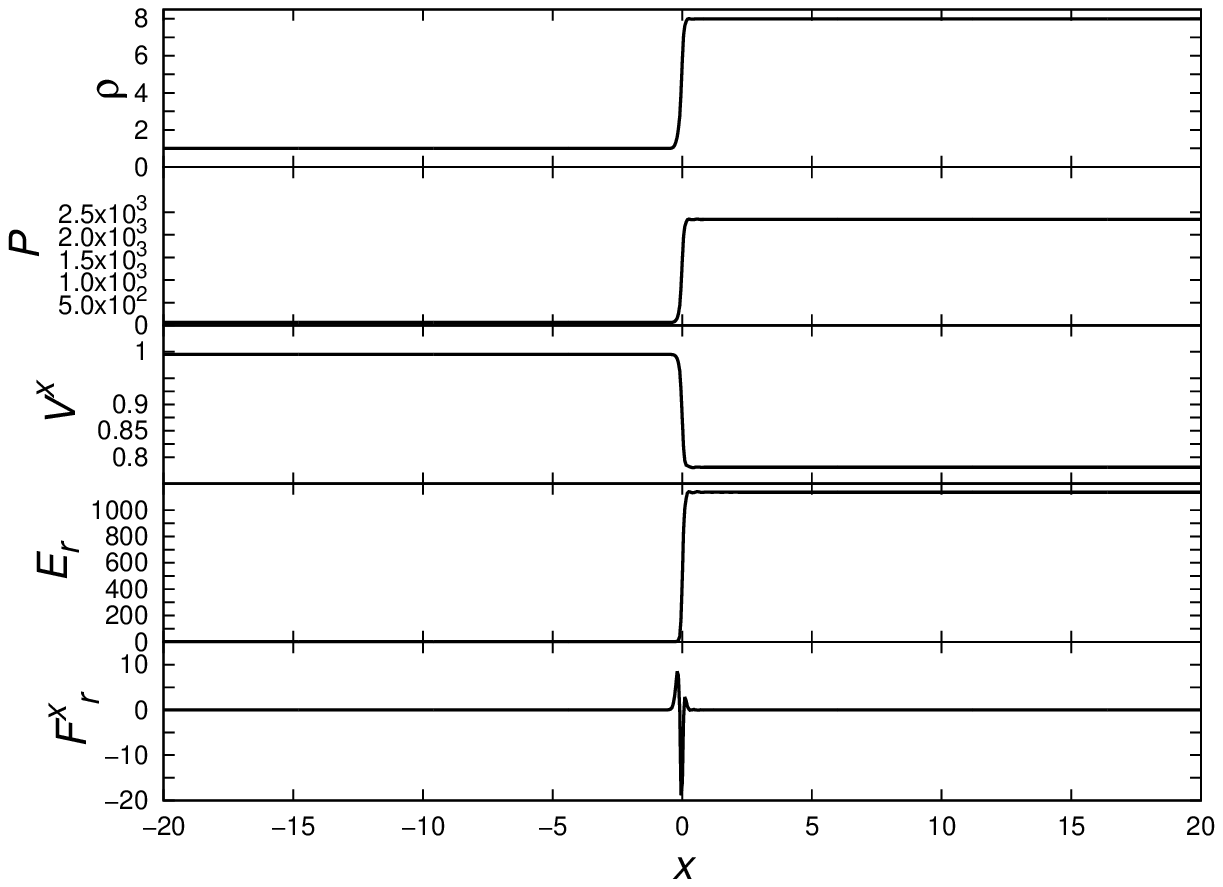}
\caption{\label{fig:t3} Tests 3. Solution of the shock tube at $t=100$ for test 3a (top) and test 3b (bottom). The panels show the rest-mass density, pressure, velocity, radiation energy density and radiation flux of the system.}
\end{figure}

(iii) \textit{Highly-relativistic wave}. This case is illustrated with Tests 3a and 3b shown in Figure \ref{fig:t3}. This is a gas-pressure dominated test in a highly relativistic regime ($W=14.195$). Test 3b is the optically thick version of this case with a high value of $\chi^t$. These two tests verify that the code is able to resolve a highly relativistic wave in two different optical depth limits.

\begin{figure}
\includegraphics[width=8.5cm]{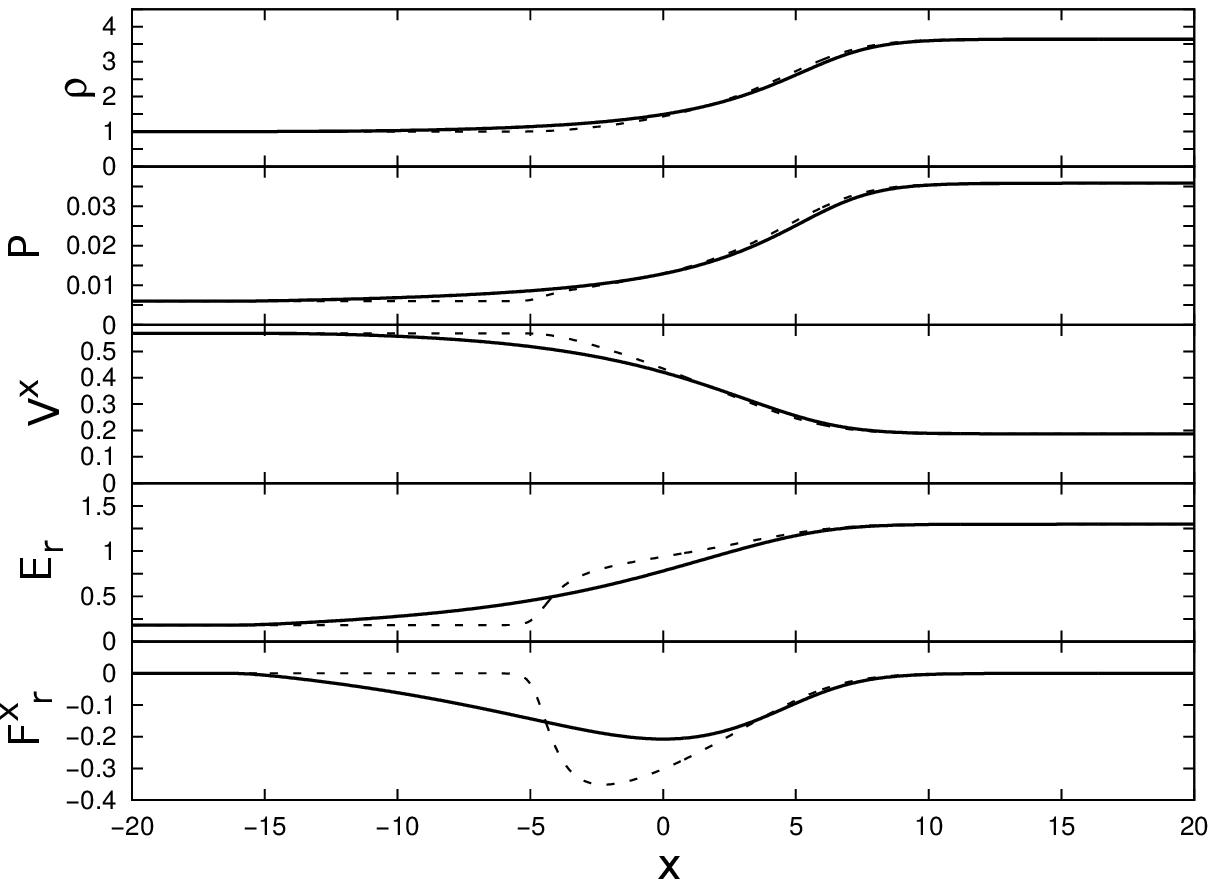}
\includegraphics[width=8.5cm]{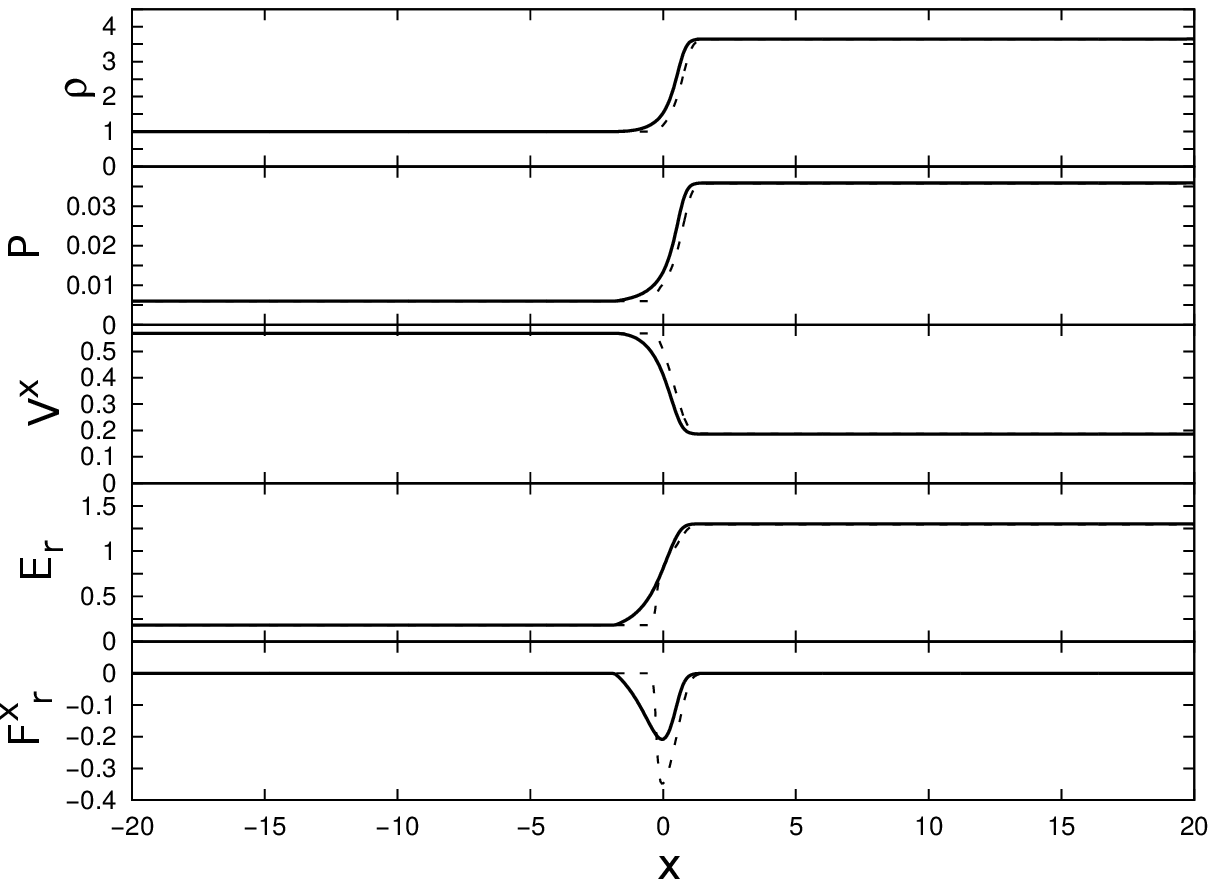}
\caption{\label{fig:t4} Tests 4. Solution at $t=500$ for test 4a (top) and 4b (bottom). The panels show the rest-mass density, pressure, velocity, radiation energy density and the radiation flux. Solid line - M1 closure, dotted line - Eddington approximation.}
\end{figure}

(iv) \textit{Radiation-pressure dominated, mildly-relativistic wave}. This case is illustrated with Tests 4a and 4b  shown in Figure \ref{fig:t4}. These tests show a shock in the radiation-dominated regime, but the shock wave is not strong and the radiation energy density in the downstream region is only a few times  higher than in the upstream region. Test $4b$ is the optically thick version of this case (Figure \ref{fig:t4} bottom). These two tests verify that the code is able to resolve a Radiation-pressure dominated wave in two different optical depth limits.

\begin{figure}
\includegraphics[width=8.5cm]{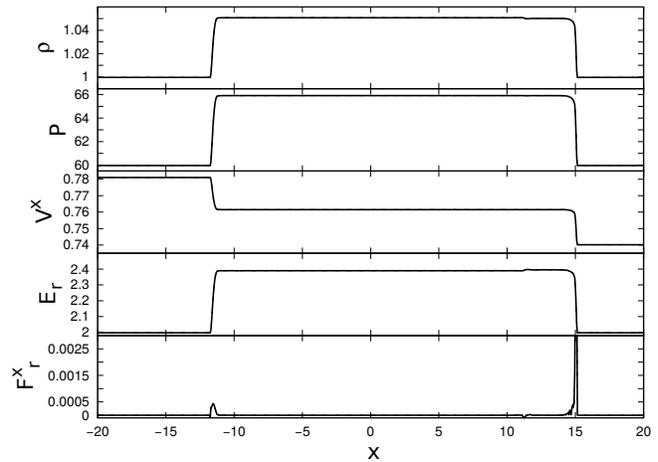}
\caption{\label{fig:t5}Snapshot  of test 5 at $t=15$. We show the rest-mass density, pressure, velocity, radiation energy density and radiation flux. Solid line - M1 closure, dotted line - Eddington approximation.}
\end{figure}

(v) \textit{Optically thick flow, mildly-relativistic}. This case is illustrated by Test 5 and is the only test that does not approach a stationary solution in the long term. The initial left and right states are identical except that they have different velocities. As a result, two shock waves propagate in opposite directions as shown during a snapshot in Figure \ref{fig:t5}.

\begin{figure}
\includegraphics[width=8.5cm]{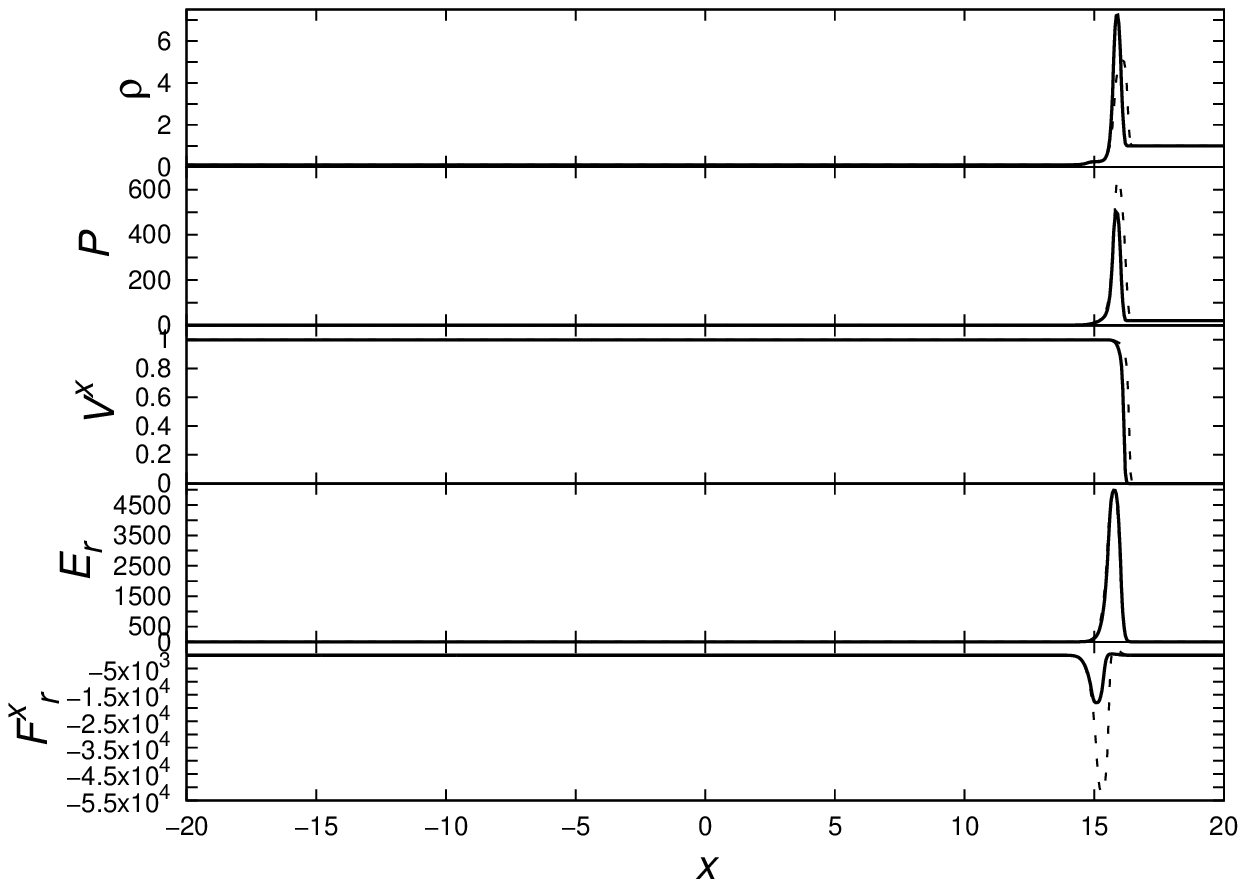}
\includegraphics[width=8.5cm]{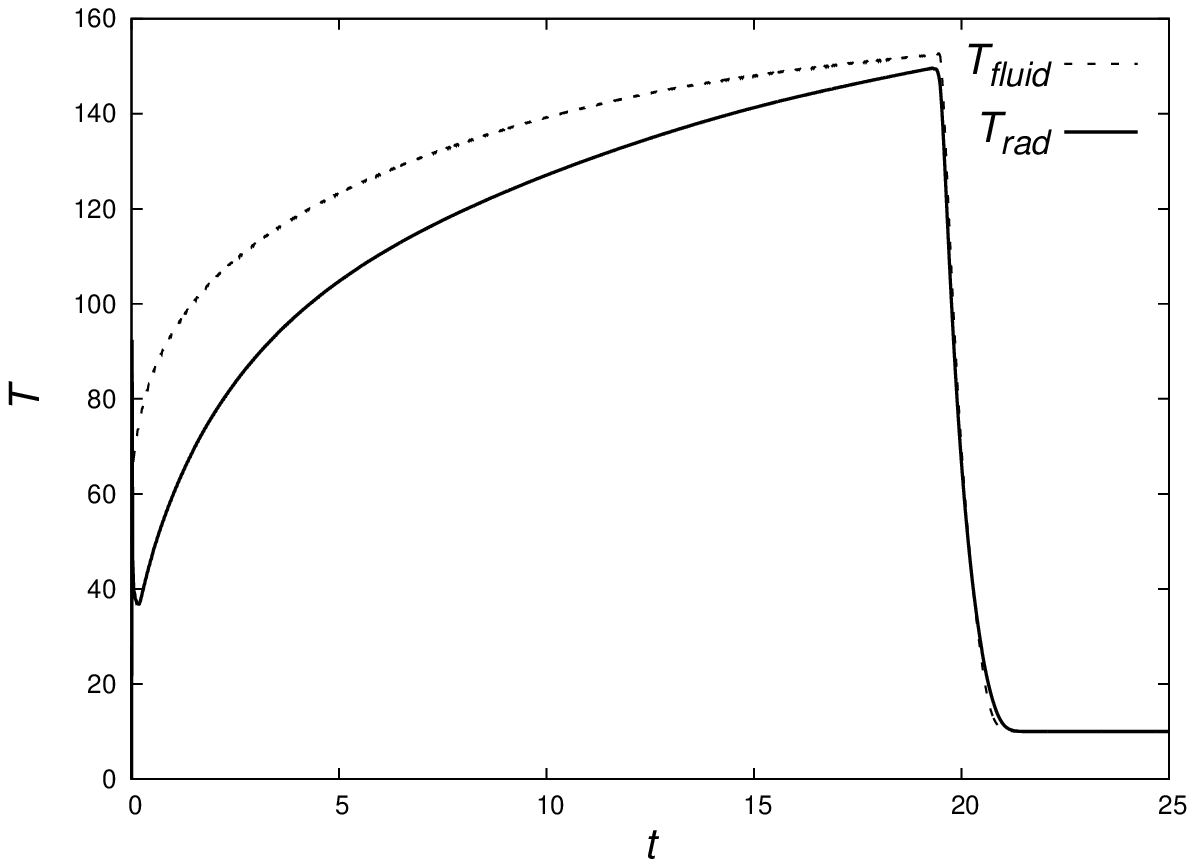}
\caption{\label{fig:t6} (Top) Snapshot  of Test 6 at $t=12$. We show the rest-mass density, pressure, velocity, radiation energy density and radiation flux. The curves with solid lines correspond to the use of the M1 closure condition, whereas the dotted line corresponds to the use of the  Eddington approximation. (Bottom) The maximum temperature in the spatial domain as a function of time when using the M1 closure. Notice that the system is clearly out of LTE during the evolution until finally both temperatures start coinciding.}
\end{figure}

(vi) {\it Shock with high Lorentz factor}. In Test 6 we show the evolution of the variables using parameters similar to those used later on to simulate GRBs. A fluid with a big Lorentz factor of the order of hundreds,  moving through a fluid at rest. We find that in a portion of the domain, there is a difference between $T_\text{fluid}$ and $T_\text{rad}$ indicating that the assumption of LTE is not valid there. However, in this case the system is driven towards thermal equilibrium at the end of evolution implying that the system relaxes, that is, there is no more energy transfer between gas and radiation. The results are found in Fig. \ref{fig:t6}.

\begin{table*}
\begin{center}
\begin{tabular}{c  c  c  c  c  c  c  c  c  c  c  c}
\hline
\hline
Test &$\Gamma$ & $a_{rad}$ & $\kappa^t$ & $\rho_L$ & $P_L$ & $u^x_L$& $E_{r,L}$ & $\rho_R$ & $P_R$ & $u^x_R$ & $E_{r,R}$ \\
\hline
 1 & 5/3 &$1.234\times10^{10}$&$0.4$&$1.0$&$3.0\times10^{-5}$&$0.015$&$1.0\times10^{-8}$&$2.4$ &$1.61\times10^{-4}$&$6.25\times10^{-3}$&$2.51\times10^{-7}$\\
\hline
 2 & 5/3 &$7.812\times10^{4}$&$0.2$ &$1.0$&$4.0\times10^{-3}$&$0.25$ &$2.0\times10^{-5}$&$3.11$&$0.04512$  &$0.0804$&$3.46\times10^{-3}$\\
\hline
 3a & 2   &$1.543\times10^{-7}$&$0.3$&$1.0$&  $60.0$          &$10.0$ &  $2.0$           &$8.0$ &$2.34\times10^{3}$&$1.25$&$1.14\times10^{3}$\\
\hline
 3b & 2   &$1.543\times10^{-7}$&$25$&$1.0$&  $60.0$          &$10.0$ &  $2.0$           &$8.0$ &$2.34\times10^{3}$&$1.25$&$1.14\times10^{3}$\\
\hline
 4a & 5/3 &$1.388\times10^{8}$&$0.08$&$1.0$ &$6.0\times10^{-3}$&$0.69$ &  $0.18$          &$3.65$&$3.59\times10^{-2}$&$0.189$ & $1.3$\\
\hline
 4b & 5/3 &$1.388\times10^{8}$&$0.7$&$1.0$ &$6.0\times10^{-3}$&$0.69$ &  $0.18$          &$3.65$&$3.59\times10^{-2}$&$0.189$ & $1.3$\\
\hline
 5 & 2   &$1.543\times10^{-7}$&$460$&$1.0$&  $60.0$          &$1.25$ &  $2.0$           &$1.0$ &$60.0$&$1.10$&$2.0$\\
\hline
 6 & 4/3   &$ 1.0\times10^{-5}$& $1.0$ & $0.1$&  $1.0$          & $999.9995$ &  $0.1$           & $1.0$ & $20$ & $0$& $0.1\times10^{-5}$\\
\hline
\hline
\end{tabular}
\caption{\label{t1} Parameters of the initial shock tube tests. $L$ stands for left state variables and $R$ for the right state variables.}
\end{center}
\end{table*}

\section{Light curves of GRBs}
\label{sec:LightCurves}

In this section, we compute the luminosity out of the radiated energy as $L=dE_r/dt$ for the following model. We assume that $X$ and gamma ray emissions are triggered by strong shocks in one dimension, and in order to investigate whether this is possible, we only need to set appropriate initial conditions similar to those for the tests we showed above. We choose such initial conditions to emulate the injection of a relativistic fluid moving through a surrounding gas at rest, which is the usual way to evolve jets. For this we define two states separated by a discontinuity at a given $x_0$, and the jet is injected from the left side of $x_0$ to the right, where we assume the surrounding medium is. To convert code units into physical units, in our code we have set $c=1$, $T=1$, $m_p=1.672\times10^{-24}M$, and $k_B=1.38\times10^{-16}M^{-1}c^{-1}$, where $M$ is the progenitor mass in cgs units. All the conversion factors between the code and cgs units are given in Table \ref{t3}.

\begin{table}
\begin{center}
\begin{tabular}{c  c}
\hline
\hline
quantity         & code to cgs units conversion factor\\
\hline
 length $x$          & $2.209\times10^{12}\left(\frac{M}{M_\odot}\right)\text{cm}$\\
 time $t$            & $73.652\left(\frac{M}{M_\odot}\right)\text{s}$\\
 velocity $v_x$      & $2.997\times10^{10}\text{cm/s}$\\
 mass density $\rho$ & $1.842\times10^{-4}\left(\frac{M_\odot}{M}\right)^2\text{g/cm}^3$\\
 energy density $E_r$ & $1.658\times10^{17}\left(\frac{M_\odot}{M}\right)^2\text{erg/cm}^3$\\
 luminosity     $L$  & $2.426\times10^{52}\text{erg/s}$\\
 opacity  $\chi$     & $2.454\times10^{-9}\left(\frac{M}{M_\odot}\right)\text{cm}^2/\text{g}$\\
\hline
\hline
\end{tabular}
\caption{\label{t3} Physical constants and conversion factors between code and cgs units. $M_\odot$ stands for the solar mass in cgs units.}
\end{center}
\end{table}

Unlike the numerical shock tests presented above, in order to approach a more realistic astrophysical scenario, in GRBs the radiation is coming from those regions in which the optical depth goes from high to low values \cite{Pe'er,Giannios}, and the external medium density $\rho_m$ and pressure profiles $P_m$ decrease with radius. To satisfy these conditions, we choose the following initial set up: (1) the initial profile of density and pressure in the surrounding region is stratified, (2) the system is optically thick, and (3) we assume local thermal equilibrium initially.

A discussion on the implications of various density and pressure profiles of the surrounding medium is a matter of current discussion \cite{Mignone,DeColle}, and in this paper we use a density and pressure profiles described by the following power law

\begin{equation}
\rho_m=\rho_0\left(\frac{x_0}{r}\right)^{2}, \ \ P_m= P_0\left(\frac{x_0}{r}\right)^{2},
\end{equation}

\noindent where the parameters of such surrounding medium are $\rho_0=n_0m_pc^2$ with $n_0= 1\text{cm}^{-3}$, and $P_0=\rho_0c^210^{-10}$.

According to  (\ref{eq:Galpha}) the system is allowed to deviate from the LTE during the evolution, and we set the condition $a_r=E_{r,b}/T_\text{fluid,b}^4$ only initially. During the evolution the temperature is calculated taking into account the contributions of the fluid and radiation pressure, that is, we solve Eq.(\ref{tem}) for $T_\text{fluid}$ using a Newton-Raphson method in the whole spatial domain and along the evolution.

Another difference between the conditions of the tests above and a more realistic astrophysical scenario involves the EoS for the gas. We use the TM EoS where the specific enthalpy is related to $P$ and $\rho$ through

\begin{equation}
h=(5/2)(P/\rho) + \sqrt{(9/4)(P/\rho)^2+1}.
\end{equation}

\noindent which is accurate in relativistic fluids \cite{Mignone}. The reason to use the TM EoS and not a gamma-law, is that the later is insufficient to distinguish cold from hot regions. Instead, the TM EoS has the property that the adiabatic index asymptotically approaches $5/3$ in the cold regions, while in the limit of high temperatures it becomes close to $4/3$ \cite{Mignone}. Regarding the radiation, we use the M1 closure because it is appropriate in the optically thin limit, a regime where radiation and matter are weakly coupled. We expect the optically thin regime to develop even though we start the evolution with optically thick conditions.

The numerical domain used is $x \in [0,400]$ with $x_0=1$. We chose a big spatial domain in order to monitor $L$ at different locations, because we expect this quantity to be dependent of the observer in the regions near the initial discontinuity and will stabilize asymptotically. We thus set 39 detectors equally spaced along the numerical domain that measure $L$.

We carried out several experiments to see how $L$ depends on the various parameters of the jet initial conditions. We produce various scenarios by changing the following parameters: the radiation energy density of the incoming beam gas $E_{r,b}$, the opacity of the gas $\chi$, the velocity of the injected gas $v_b$ that will impact in its Lorentz factor associated $W_b$, and the density of the incident gas $\rho_b$. In all cases the pressure of the incident gas and the surrounding medium are equal $P_b = P_0$ at the interface $x_0$.

\begin{figure}
\includegraphics[width=8.9cm]{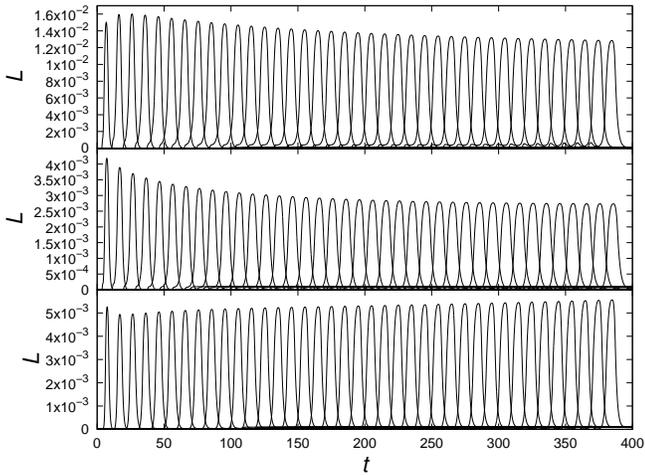}
\caption{\label{det} Light curves measured by various detectors located along the domain in code units. The parameters used in these curves from top to bottom are: 
(Top) $W_b=800,~\chi^t=1,~E_{r,b}=1\times 10^{-5},~\rho_b =1$,
(Middle) $W_b=600,~\chi^t=0.1,~E_{r,b}=1\times 10^{-5},~\rho_b =1$,
(Bottom) $W_b=800,~\chi^t=10,~E_{r,b}=1\times 10^{-3},~\rho_b =0.1$.
Notice that high Lorentz factors produce higher signals. Taking in mind that $x_0=1$, and then, detectors on the left measure a significantly different signal than those far from the initial discontinuity. We choose the signal measured by the farthest detector for our analyses.}
\end{figure}

In Figure \ref{det} we show the light curves in code units measured by various detectors for three particular cases with different Lorentz factor, opacity, radiated energy, and rest-mass density of the injected gas. The detectors are located at $x_d=10,20,... 400$. The curves on the left are the signals measured by detectors located at the left of the domain near $x_0$, whereas those to the right are the signals measured by the detectors on the right part of the domain, far from $x_0$. It can be seen that the signals are considerably different in regions near the initial shock whereas the signal becomes more detector-independent far from the initial shock. Based on the detector dependence of the light curve, we will consider the signal observed in an asymptotic region will be the signal measured by the farthest detector located at $x_d=400$. 

In Table \ref{t2} we show the parameters for different cases, each one evolved for three different values of $W_b$ and $\rho_b$. Notice that the initial value for the radiated energy density in the outer medium $E_{r,m}$ is not given in this Table, but this value is calculated imposing the local thermal equilibrium on the right state, this is $E_{r,m}=a_{r}T_\text{fluid,m}^{4}$. In the left part of Table \ref{t2} we show nine experiments within the gray-body approximation. On the right side of Table \ref{t2} we present other nine cases using opacities corresponding to Thomson scatettering and Bremsstrahlung radiation.

\begin{table*}
\begin{center}
\begin{tabular}{c c c c c|| c c c c c }
\hline
\hline
\multicolumn{5}{c||}{Gray-Body }  & \multicolumn{5}{c}{Thomson scatettering and Bremsstrahlung}\\
\hline
\# \ \ & $\rho_b \ (\text{g}/\text{cm}^3)$ \ \  & $W_b$ \ \ & $E_{r,b} \ (\text{erg/cm}^3)$ \ \ & $\chi^t \ (\text{cm}^2/\text{g})$ & \# &$\rho_b \ (\text{g}/\text{cm}^3)$ \ \  & $W_b$ \ \ & $E_{r,b} \ (\text{erg/cm}^3)$  \\
\hline
$1$ \ \ & $\rho_0$ \ \ & $800$ \ \ & $1.658\times10^{11}$ & $0.245\times10^{-8}$ & $10$ &$\rho_0/10$ \ \ & $800$ \ \ & $1.658\times10^{13}$\\
$2$ \ \ & & & &$2.454\times10^{-8}$ &$11$ & & &$1.658\times10^{11}$\\
$3$ \ \ & & & &$24.54\times10^{-8}$ &$12$ & & &$1.658\times10^{10}$\\
$4$ \ \ & $\rho_0$ \ \ & $600$ \ \ &$1.658\times10^{9}$ & $0.245\times10^{-8}$ &$13$& $\rho_0$  \ \    & $600$ \ \ & $1.658\times10^{13}$\\
$5$ \ \ & & & &$2.454\times10^{-8}$ &$14$ & & &$1.658\times10^{11}$\\
$6$ \ \ & & & &$24.54\times10^{-8}$ &$15$ & & &$1.658\times10^{10}$\\
$7$ \ \ & $\rho_0/10$ \ \ & $200$ \ \ & $1.658\times10^{11}$ & $0.245\times10^{-8}$ &$16$& $\rho_0$ \ \  & $200$ \ \ & $1.658\times10^{13}$\\
$8$ \ \ & & & &$2.454\times10^{-8}$ & $17$ & & &$1.658\times10^{11}$\\
$9$ \ \ & & & &$24.54\times10^{-8}$ &$18$ & & &$1.658\times10^{10}$\\
\hline
\hline
\end{tabular}
\caption{\label{t2} This table contains the initial parameters of jets for 18 different numerical experiments that will tell us the effects of each parameter on the shape of the final light curve. The left part shows parameters for gray-body experiments where $\chi^s=0$. The right part shows the initial parameters  considering opacities for Thomson scattering and Bremsstrahlung radiation. All these experiments were made for $M=10M_\odot$. In all the experiments we use the following parameters for the medium: $P_b = P_0$, $W_m=0.0$.}
\end{center}
\end{table*}

\begin{figure*}
\centering
\includegraphics[width=8.cm]{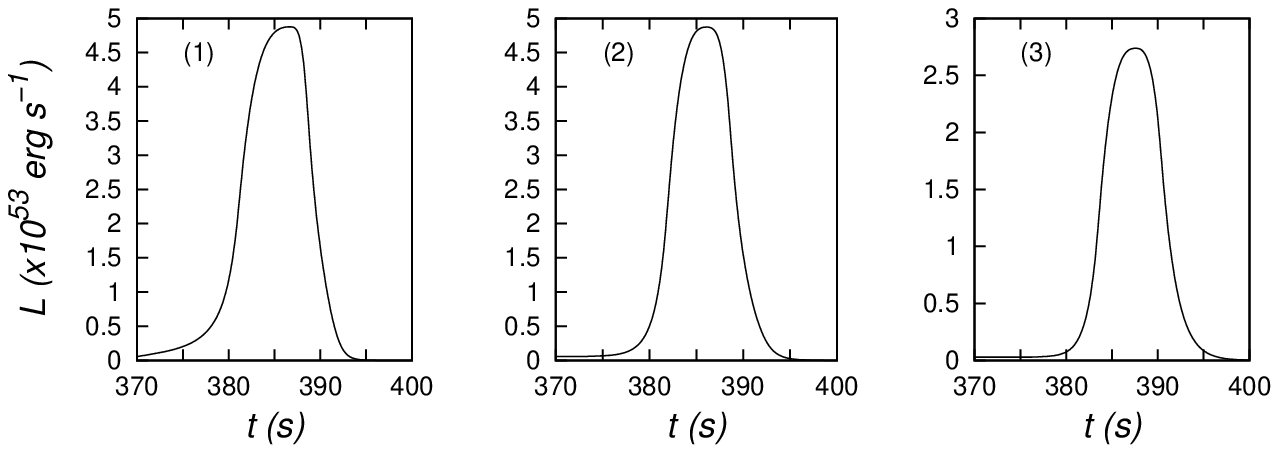}
\includegraphics[width=8cm]{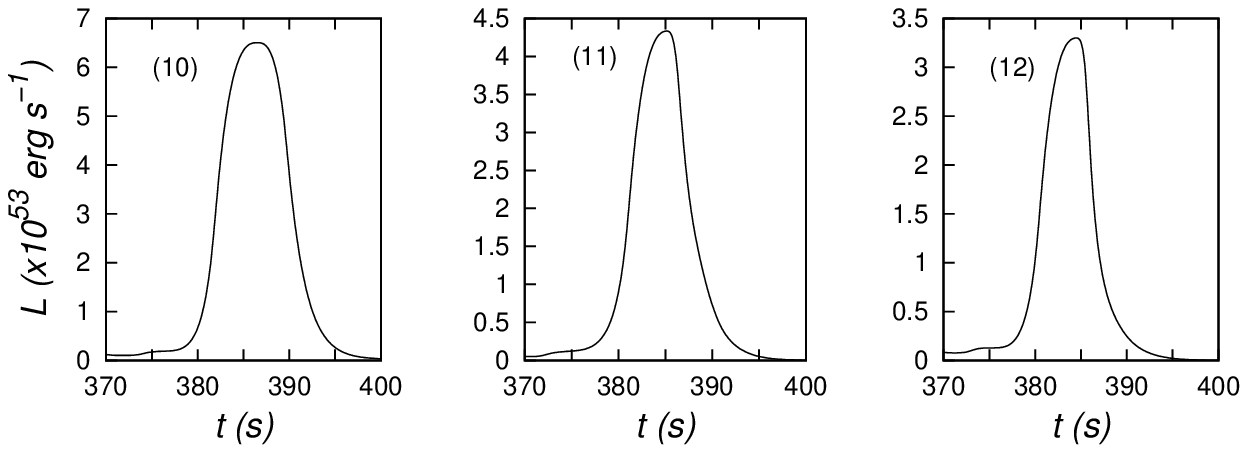}
\includegraphics[width=8cm]{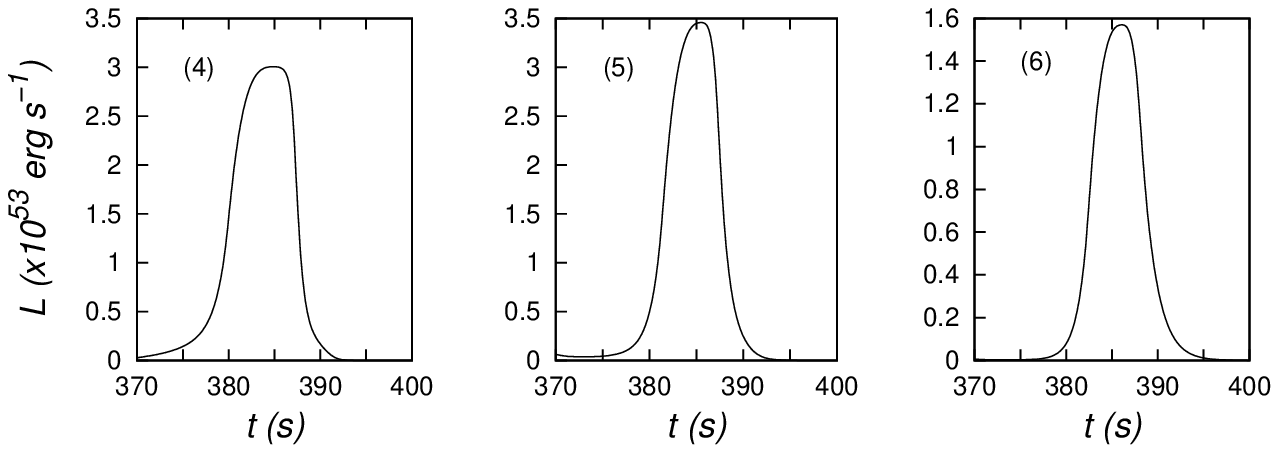}
\includegraphics[width=8cm]{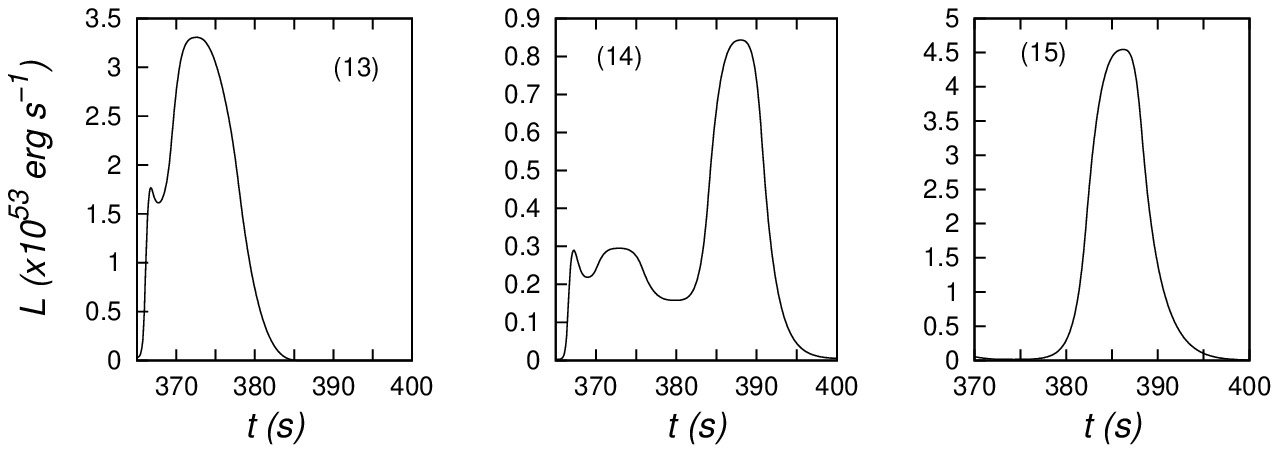}
\includegraphics[width=8cm]{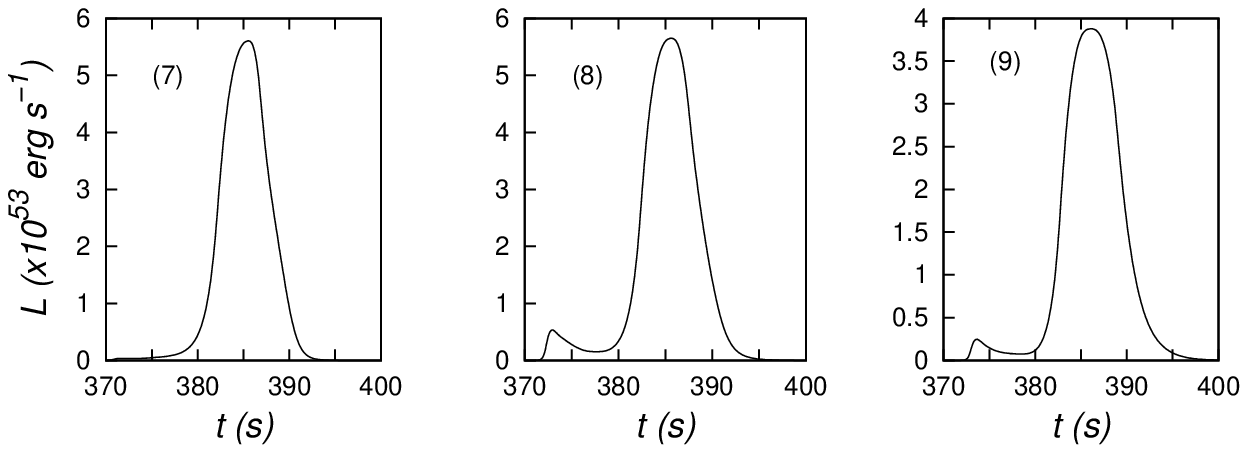}
\includegraphics[width=8cm]{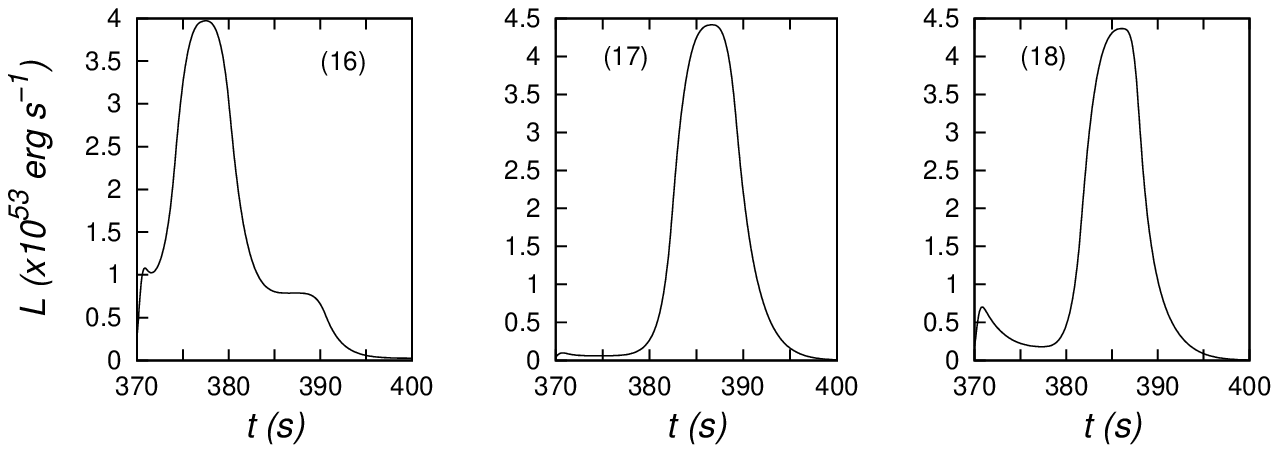}
\caption{\label{fig:catalog} Sample of our catalog with different types of light curves depending on the various combinations of the initial state parameters described in Table \ref{t2}. The nine experiments on the left correspond to the nine cases assuming the gray body approximation. Those on the right correspond to Thomson scattering and Bremsstrahlung radiation.}
\end{figure*}

In the left side of Fig. \ref{fig:catalog} we show the impact of changing the various parameters of the injected gas in the signal of the farthest detector, considering only the effects of the thermal opacity from the gray-body.  For instance, it can be seen that all curves show a sharp high peak like that of GRBs.

Next we consider a different type of experiments, in this case using opacities corresponding to Thomson scattering and Bremsstrahlung opacities, these opacities are \cite{George}.

\begin{eqnarray}
\kappa^s &=& 0.4\rho^2_{\text{cgs}} \text{cm}^{-1}, \nonumber \\
\kappa &=& 1.7\times10^{-25}T_K^{-7/2}\rho^2_{\text{cgs}} m_p^{-2}\text{cm}^{-1}, \nonumber
\end{eqnarray}

\noindent where $T_K$, $\rho_{\text{cgs}}$, and $m_p$ are the temperature of the fluid in Kelvin, the density and the mass of a proton, respectively in cgs units. Bremsstrahlung is the main cooling process at temperatures $T>10^7$K and the optically thin limit. This process is considered important in the afterglow phase, because any internally generated radiation can freely escape from the emitting region in a band of energy of KeV \cite{George}. In an optically thin region, this is a better approximation than a gray body. In the regime where Bremsstrahlung radiation dominates, the primary source of opacity is the Thomson scattering. In the right part of Fig. \ref{fig:catalog} we show the impact of the opacity for Thomson scattering and Bremsstrahlung radiation  in the signal measured by the farthest detector for three different values of radiated energy.

In Fig. \ref{fig:catalog} we thus show the catalog of light curves calculated for each of the parameters in Table \ref{t2}. These plots have similar properties to those of light curves of GRBs, that is, a prompt emission phase followed by an afterglow phase. Besides, in this catalog we can see that light curves have luminosities of order $10^{52} - 10^{53} erg~ s^{-1}$, which is a safe range within  the observational evidence indicating that light curves of GRBs have a magnitude between $10^{51}-10^{54} erg~ s^{-1}$ \cite{Pescalli}. These experiments serve to bound the parameters that are compatible with the observations. In the experiments of Table \ref{t2} we have a  combination of the all parameters involved. Now we can summarize qualitatively the individual effects of each of these parameters on the light curves. In these experiments we assume the progenitor mass to be $M=10M_\odot$, so that it would correspond to a stellar mass black hole progenitor \cite{Delaurentis}.

\begin{enumerate}

\item \textit{Effects of the Lorentz factor.} The impact of this parameter is reflected in the width and amplitude of the light curves. In the case of the gray body approximation, the light curves are characterized by a single pulse. The bigger the Lorentz factor the bigger the duration and amplitude of the pulse. In the case of Thomson scattering and Bremsstrahlung radiation, the light curves exhibit diferent morphologies, if the Lorentz factor of the jet is of $800$, the light curves also are characterized by a single pulse. When the Lorentz factor is smaller, the light curves are wider and can exhibit one peaked signal together with other pulses of smaller amplitude.

\item \textit{Effects of opacity.} This parameter can be considered the most important of the four free parameters, because contains the information about the radiative processes happening during the evolution. In the case of Thomson scattering and radiation Bremsstrahlung, the light curves exhibit different morphologies, if the Lorentz factor is of around 800, the light curves only show one peak followed by a smooth decay, if the Lorentz factor in less than 800, the curves exhibit various small peaks  before or after the main peak. In the case of the gray body approximation, there is not a variation in morphology of the light curves if the Lorentz factor is of order 800, if the Lorentz factor is 200 the light curves show a small signal before the prompt emission. In both cases the amplitude increases when the Lorentz factor does.

\item \textit{Effects of the injected gas density.} All experiments indicate that this parameter has an impact analogous to the Lorentz factor. But unlike the Lorentz factor, the mass density of the injected gas has an inverse effect, that is, when $\rho_b$ is an order of magnitude smaller than the magnitude of the  surrounding medium the amplitude increases even $25 \%$ and has a width of around 10 seconds smaller compared to the width when both densities are equal.

\item \textit{Effects of radiated energy density.} This parameter significantly affects the amplitude of the light curves whereas the width remains unchanged. The bigger the radiation energy injected with the jet, the bigger the luminosity peak of the curves. This parameter has the same effect that the Lorentz factor considered, that is, the light curves are wider and have higher amplitude when $E_{r,b}$ is bigger.

\end{enumerate}

A simple descriptions in general terms of these experiments indicates that all the light curves shown in the gray body approximation have the same morphology in all experiments, that is, a single pulse followed by a smooth decay. In the Thomson scattering cases, there is a slight knee previous to the peak.

\subsection{Particular GRBs}

Now, we apply the knowledge learned in the experiments to observed GRB light curves. For this we need to convert the luminosity $L$ to radiated flux $F_r$, using the luminosity distance $d_L$  for a steady, isotropically emitting source given by the expression $F_r=L/4\pi d_L^2$ \cite{George}.

We  choose two long GRBs on an energy band $15-150$ KeV taken from \cite{Mendoza}, observed by BAT on board Swift satellite. These are GRB080413B and GRB060904B. The initial conditions for the jet and surrounding medium that best fit the light curves are

\begin{itemize}
\item[-]  
For GRB080413B, $W_b=645.497$, $E_{r,b}=5.158\times10^{8} erg/cm^3$, $\rho_b=\rho_0$, and $W_m=0$. For this fit, we use a progenitor mass of $M=15M_\odot$. 

\item[-] 
For GRB060904B, $W_b=674.2$, $E_{r,b} = 2.155\times10^{10} erg/cm^3$, $\rho_b = \rho_0$, and $W_m=0$. For this fit, we use a progenitor mass of $M=10M_\odot$. 

\end{itemize}

\begin{figure}
\includegraphics[width=8.9cm]{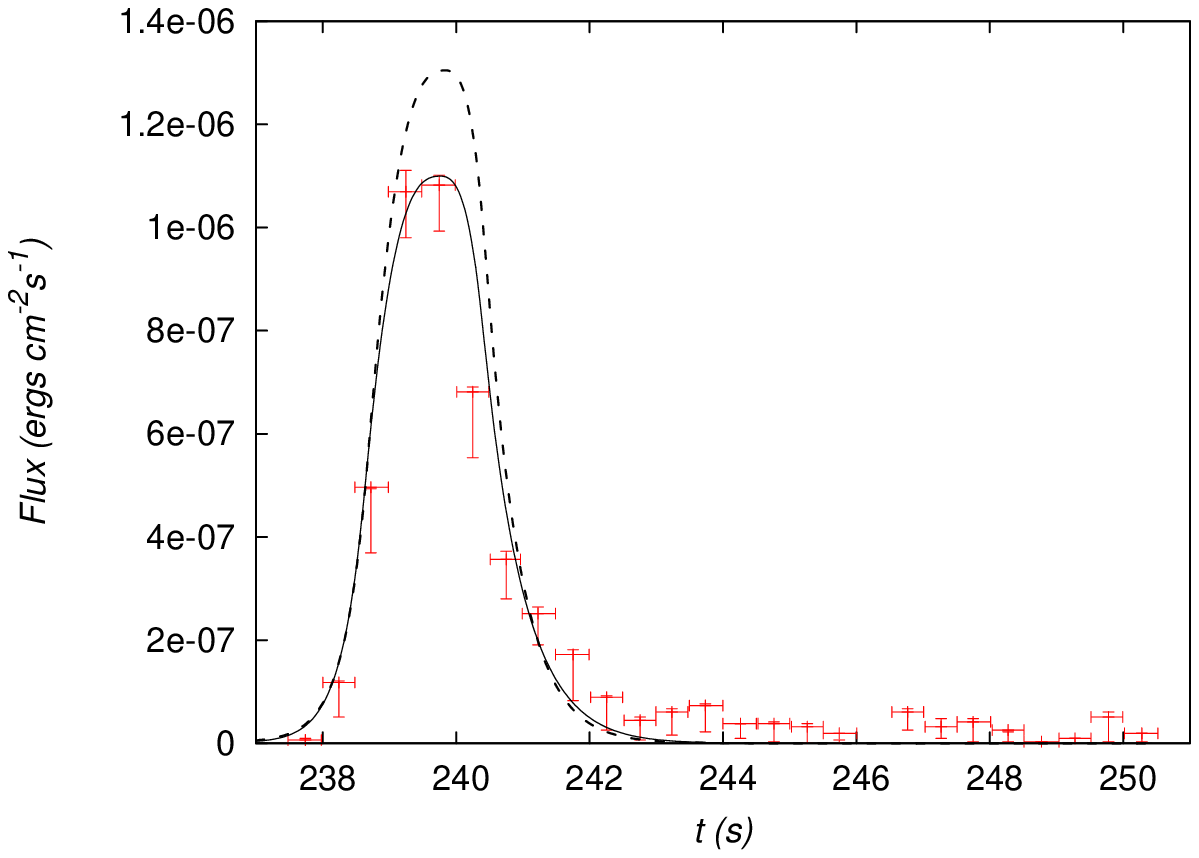}
\includegraphics[width=8.9cm]{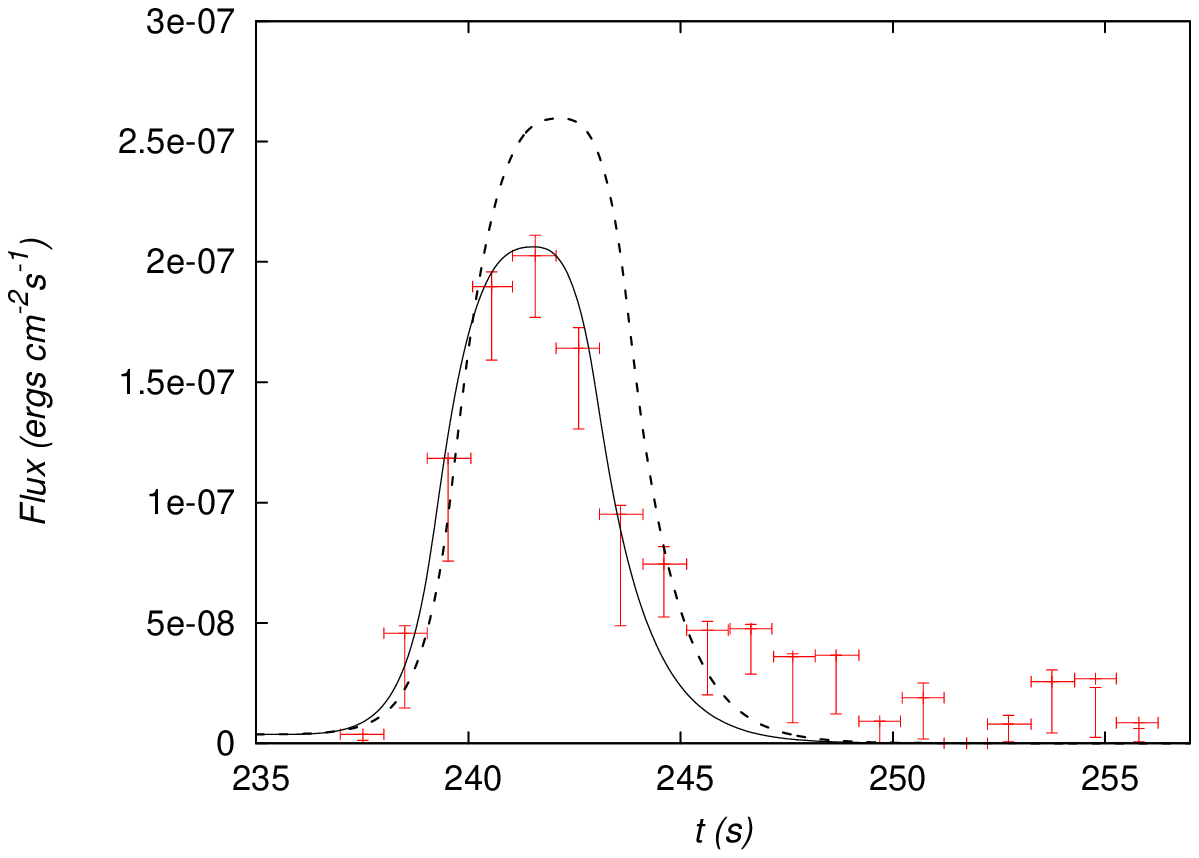}
\caption{\label{GRBs} The light curve of two long GRBs GRB080413B (top) and GRB060904B (bottom), the luminosity peak for these GRBs are $1.49\pm 0.18\times 10^{52}$ and $2.042 \times 10^{52}$ erg/s respectively. The data represented by dots with error bars were taken from (Mendoza et al. 2009). The solid lines correspond to the opacities given by Thomson scattering and Bremsstrahlung radiation, and dotted lines to the gray body model.}
\end{figure}

In Fig. \ref{GRBs} we show the fits for these two GRBs. The fit for these GRBs were made taking into account the cases where the opacities given by Thomson scattering and Bremsstrahlung radiation (solid line), and thermal opacity (dotted line). The value for thermal opacity is $\chi^t = 3.313\times10^{-9} cm^2/g$ and $\chi^t = 1.3\times10^{-10} cm^2/g$ for GRB080413B and GRB060904B, respectively. When we are considering Thomson scattering and Bremsstrahlung radiation we obtain a better fit than when using the gray body approximation. However none of the two models fits well the afterglow phase.

\begin{figure}
\includegraphics[width=8.9cm]{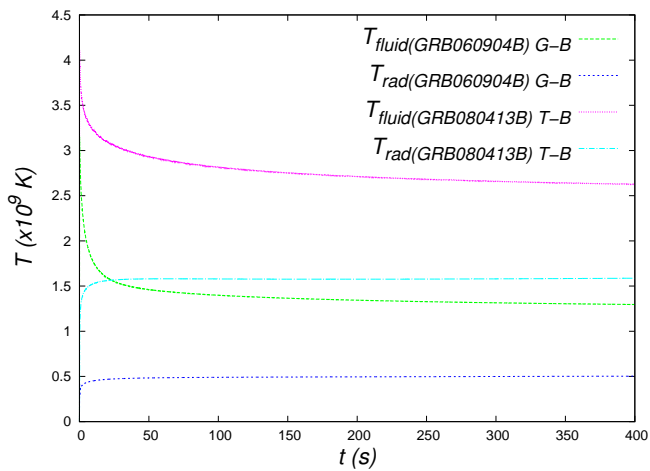}
\caption{\label{GRBsT} Comparison of the profiles of maximum temperature of the fluid and radiation for the gray body approximation (G-B), and Thomson scattering and Bremsstrahlung radiation (T-B).}
\end{figure}

In order to show how out of LTE the fluid and radiation are during the process, in Fig. \ref{GRBsT} we show the maximum temperature of the fluid and radiation in time. The temperatures associated with GRB060904B in the gray body approximation are $T_{fluid(060904B)G-B}$ and $T_{rad(060904B)G-B}$, whereas the temperatures associated with GRB080413B in the Thomson scattering and Bremsstrahlung radiation are $T_{fluid(080413B)T-B}$ and $T_{rad(080413B)T-B}$. We can say that in both models the fluid and radiation temperatures are significantly different. In the two models and two cases the maximum temperatures behave as $T_\text{fluid}>T_\text{rad}$. This suggests, according to eq. (\ref{eq:Galpha}), that the fluid cools down during the process.

In our fits we calculate the chi-square associated with the fit, using $\chi^2 = \sum_{i=1}^{N}\left(\frac{m_{o,i}-m_{t,i}}{\sigma_i}\right)^2$, where, $m_o$ is the apparent magnitude observed, $m_t$ is the apparent magnitude theoretical, $\sigma_i$ is the standard deviation of the observations, and $N$ is the the total number of available data. Next, we show the values of reduced chi-squared for the best fit of GRBs above, which is defined as $\chi^2/dof$ where $dof$ stands for degrees of freedom. These fits correspond to the case Thomson scattering and Bremsstrahlung radiation, and the values are  $\chi_{08}^{2}/dof = 4.3075$ and $\chi_{06}^{2}/dof = 2.5027$ for GRB080413B and GRB060904B, respectively.

\section{Discussion and conclusions}
\label{sec:conclusions}

We have solved the 1D RRH equations to calculate the light curves produced by initial conditions of jets with the aim to investigate whether a single shock can explain long GRBs. Numerical simulations were made considering four free parameters, and under the initial assumption of local themal equilibrium between the fluid and the radiation field. The free parameters are: rest mass density, Lorentz factor, radiated energy, and opacity. We constructed the light curves for many combinations of these parameters and fit  the light curves of two particular GRB light curves.

Regarding the radiation field, we have considered two models. In the first one the thermal opacity is  proportional to the rest mass density and the scattering opacity is zero. In the second model the opacity is given by Thomson scattering and Bremsstrahlung radiation. There is a significant difference between these models. In the first one, we find that light curves for $W_b>200$ have the same morphology, that is, a single pulse followed by smooth decay, whereas for $W_b\sim 200$ and $\chi^t \sim10^{-7}cm^2/gr$ we found that the light curves show the burst preceded by small pulses. In the second model we find three different morphologies: the first scenario with a single pulse, the second shows light curves that have one pulse followed by a knee and the third scenario is characterised by light curves with one or various pulses before the main pulse followed by a smooth although fast decay.

As a first contact with observations, we fit  the light curves associated with two long GRBs, GRB080413B and GRB060904B. The model that best fits the light curves uses an opacity emulating Thomson scattering and Bremsstrahlung radiation, which describes well the prompt emission, but is not enough to explain the afterglow. The best fits of the two GRBs correspond to continuous fluid density $\rho_b=\rho_0$ at $x_0$. The Lorentz factor of the incident beam is $W_b=645.497.6$ for GRB080413B and  $W_b=674.2$ for GRB060904B.

This model seems to do well at explaining the GRB peak of luminosity curves, however it is not as good in the afterglow phase. In order to determine whether this single shock model can improve, we will explore a wider sample of long GRBs and various surrounding density profiles \cite{DeColle}. We will also add multifrequency analysis similar as done in \cite{Cuesta}.



\section*{Acknowledgments}

This research is partly supported by grants CONACyT 258726 and CIC-UMSNH-4.9.



\begin{thebibliography}{99}
\bibitem[Cuesta-Mart\'inez, Aloy \& Mimica 2015]{Cuesta}
Cuesta-Mart\'inez C., Aloy M. A., Mimica P., 2015, MNRAS, 446, 1716-1736.
\bibitem[Cuesta-Mart\'inez et al. 2015]{CuestaII}
Cuesta-Mart\'inez C., Aloy M. A., Mimica P., Thone C., de Ugarte Postigo A., 2015, MNRAS, 446, 1737-1749.
\bibitem[De Colle et al. 2012]{DeColle}
De Colle F., Ramirez-Ruiz E., Granot J., Lopez-Camara D., 2012, ApJ 751, 57.
\bibitem[De Laurentis et al. 2015]{Delaurentis}
De Laurentis M., Garufi F., Dainotti M. G., Milano L., 2015, arXiv:1506.00106.
\bibitem[Dermer \& Mitman 1999]{Dermer}
Dermer C.D., Mitman K.E., 1999, ApJ, 513, L5.
\bibitem[Dubroca \& Feugeas 1999]{Dubroca}
Dubroca B., Feugeas J. L. 1999, CRAS, 329, 915.
\bibitem[Farris et al. 2008]{Farris}
Farris B. D., Li T. K., Liu Y. T., Shapiro S. L., 2008, Physical Review D, 78, 024023.
\bibitem[Fragile et al. 2012]{Fragile}
Fragile P. C., Gillespie A., Monahan T., Rodriguez M., Anninos P., 2012, ApJS, 201, 9. 
\bibitem[Gonz\'alez, Audit \& Huynh 2007]{Gonzalez}
Gonz\'alez M., Audit E., Huynh P., 2007, A\&A, 464, 429.
\bibitem[Pe'er \& Ryde 2011]{Pe'er}
Pe'er A. \& Ryde F., 2011, ApJ 732, 49.
\bibitem[Giannios 2012]{Giannios}
Giannios D.,2012, MNRAS 422, 3092.
\bibitem[Kobayashi, Piran \& Sari 1997]{Kobayashi}
Kobayashi S., Piran T., Sari R., 1997, ApJ, 490, 92.
\bibitem[Kumar \& Zhang 2015]{kumar}
Kumar P., Zhang B., 2015, Phys. Rep., 561, 1.
\bibitem[Levermore 1984]{Levermore}
Levermore C. D., 1984, J. Quant. Spectrosc. Radiative Transfer, 31, 149.
\bibitem[Lyutikov \& Blandford 2003]{Lyutikov}
Lyutikov M., Blandford R., 2003, arXiv:astro-ph/0312347. 
\bibitem[Mendoza et al. 2009]{Mendoza}
Mendoza S., Hidalgo J. C., Olvera D., Cabrera J. I., 2009, MNRAS, 395, 1403.
\bibitem[Mignone et al. 2005]{Mignone}
Mignone A., Plewa T., Bodo G., 2005, 2005 ApJS 160, 199.
\bibitem[Mihalas \& Mihalas 1984]{MihalasII} 
Mihalas D., Mihalas W., 1984, Foundations of Radiation Hydrodynamics, Oxford University Press.
\bibitem[Mimica et al. 2009]{Mimica}
Mimica P., Aloy M. A., Agudo I., Mart\'in J. M., G\'omez J. L., Miralles J. A., 2009, ApJ, 696, 1142.
\bibitem[Paczynski 1998]{Paczynski}
Paczynski B., 1998, ApJ, 494:L45.
\bibitem[Pareschi \& Russo 2005]{Pareschi}
Pareschi L., Russo G., 2005, Journal of Scientific Computing, 25.
\bibitem[Pescalli et al. 2015]{Pescalli}
Pescalli A., Ghirlanda G., Salvaterra R., Ghisellini G., Vergani S. D., Nappo F., Salafia O. S., Melandri A., Covino S., Gotz D., 2016, A\&A 587, A40.
\bibitem[Rees \& M\'esz\'aros 1992]{Rees}
Rees M. J., M\'esz\'aros P., 1992, MNRAS, 258, 41.
\bibitem[Roedig, Zanotti \& Alic 2012]{ZanottiII}
Roedig C., Zanotti O., Alic D. 2012, MNRAS, 426, 1613.
\bibitem[Roger 1977]{Alexander}
Roger A., 1977, SIAM Journal on Numerical Analysis, Vol. 14, No. 6, 1006. 
\bibitem[Rybicky \& Lightman  2004]{George}
Rybicky G. B., Lightman A. P., 2004, Radiative processes in astrophysics, WILEY-VCH.
\bibitem[Sadowski et al. 2013]{Sadowski} 
Sadowski A., Narayan R., Tchekhovskoy A.,  Zhu Y., 2013, MNRAS, 429, 3533.
\bibitem[Takahashi et al. 2013]{Takahashi}
Takahashi H. R., Ohsuga K., Sekiguchi Y., Inoue T., Tomida K., 2013, ApJ, 764, 122.
\bibitem[Tolstov et al. 2014]{Tolstov}
Tolstov A., Blinnikov S., Nagataki S., Nomoto K., 2015, ApJ, 811, 47.
\bibitem[Uri, Steven \& Raymond 1997]{Uri}
Uri M. A., Steven J. R., Raymond J. S., 1997, Applied Numerical Mathematics, 25, 151.
\bibitem[Woosley 1993]{Woosley}
Woosley S. E, 1993, ApJ 405, 273.
\bibitem[Zanotti et al. 2011]{ZanottiI}
Zanotti O., Roedig C., Rezzolla L., Del Zanna L., 2011, MNRAS, 417, 2899.
\bibitem[Zhang \& Yan 2011]{Zhang}
Zhang B., Yan H., 2011, ApJ 726, 90.
\end{thebibliography}
\end{document}